\documentclass[12pt]{article}
\usepackage{graphicx}
\newcommand {\nc} {\newcommand}
\nc {\beq} {\begin{eqnarray}}
\nc {\eeq} {\end{eqnarray}}
\nc {\eeqn} [1] {\label{#1} \end{eqnarray}}
\nc {\eoln} [1] {\label{#1} \\}
\nc {\eol} {\nonumber \\}
\nc {\rref} [1] {(\ref{#1})}
\nc {\Eq} [1] {Eq.~(\ref{#1})}
\nc {\Ref} [1] {Ref.~\cite{#1}}
\nc {\la} {\mbox{$\langle$}}
\nc {\ra} {\mbox{$\rangle$}}
\nc {\cL} {\mbox{${\cal L}$}}
\nc {\dem} {\mbox{$\frac{1}{2}$}}
\nc {\ve} [1] {\mbox{\boldmath $#1$}}
\nc {\arrow} [2] {\mbox{$\mathop{\rightarrow}\limits_{#1 \rightarrow #2}$}}
\begin{document}
\title{Confined helium on Lagrange meshes}
\author{D. Baye$^a$ and J. Dohet-Eraly$^b$ \\
$^a$ Physique Quantique, and \\ Physique Nucl\'eaire Th\'eorique et Physique Math\'ematique, C.P. 229, \\
Universit\'e Libre de Bruxelles (ULB), B-1050 Brussels Belgium \\
$^b$ TRIUMF, 4004 Wesbrook Mall, Vancouver, \\ British Columbia V6T 2A3, Canada}  
\maketitle
\begin{abstract}
The Lagrange-mesh method has the simplicity of a calculation on a mesh 
and can have the accuracy of a variational method. 
It is applied to the study of a confined helium atom. 
Two types of confinement are considered. 
Soft confinements by potentials are studied in perimetric coordinates. 
Hard confinement in impenetrable spherical cavities is studied in a system 
of rescaled perimetric coordinates varying in [0,1] intervals. 
Energies and mean values of the distances between electrons and 
between an electron and the helium nucleus are calculated. 
A high accuracy of 11 to 15 significant figures is obtained 
with small computing times. 
Pressures acting on the confined atom are also computed. 
For sphere radii smaller than 1, their relative accuracies are better than $10^{-10}$. 
For larger radii up to 10, they progressively decrease to $10^{-3}$, 
still improving the best literature results. 
\end{abstract}
\section{Introduction}
\label{sec:intro}
Many numerical techniques exist for quantum-mechanical calculations 
in configuration space. 
Among them, two main qualities may be searched: accuracy and simplicity. 
However, they are not often encountered simultaneously. 
For some problems, the Lagrange-mesh method has the accuracy of a variational method 
and the simplicity of a calculation on a mesh \cite{BH86,VMB93,BHV02,Ba06,Ba15}. 
This approximate variational method involves a basis of Lagrange functions, 
i.e.\ infinitely differentiable functions vanishing at all mesh points 
of a Gauss quadrature, except one. 
With the help of the associated Gauss quadrature, all matrix elements are very simple. 
In particular, the matrix elements of the potential are approximated 
by values of the potential at mesh points like in collocation methods. 

The striking property of the Lagrange-mesh method is that, in spite of its simplicity, 
it has essentially the same accuracy as a variational calculation performed 
with the same Lagrange basis. 
This performance is not well understood yet \cite{BHV02}. 
However, the accuracy of the method depends on the validity of the Gauss quadrature. 
Hence, the method can be very bad in the presence of singularities of the potential. 
This problem can sometimes be cured by a so-called regularization \cite{VMB93,BHV02,Ba06,Ba15}. 
The method has been successfully applied to many problems in atomic, molecular and nuclear 
physics (see \Ref{Ba15} for a review). 
It is particularly useful to solve coupled-channel problems in the continuum \cite{HSV98,DB10} 
or three-body problems \cite{HB03,OB12}. 

The aim of the present paper is to apply the Lagrange-mesh method to a three-body problem: 
a helium atom confined in some environment \cite{AFR03,AGF06,LC09,FAM10,MAF10,WMS10,BSM13}. 
The confinement can simulate a helium gas under pressure or helium atoms 
trapped in some molecule, carbon cluster or crystal. 
As we show below, existing techniques \cite{HB99} are very convenient for soft 
confinements by potentials but can not be applied for an atom confined 
in an impenetrable spherical cavity. 
The difficulty arises from simultaneously meeting two different constraints: 
regularizing the singularities of the three Coulomb terms and forcing the confinement. 
The former condition is easily treated in perimetric coordinates \cite{CJ37,Pe58} 
which automatically regularize the singularities but which necessarily extend 
over the whole configuration space. 
Confinement is easily and accurately treated on a Lagrange mesh 
over a finite interval for hydrogen \cite{BS08,Ba15}. 
Here we show that a new coordinate system can be built for helium which keeps the regularization 
of the potential but over an impenetrable spherical cavity. 

The principle of the Lagrange-mesh method is recalled in section \ref{sec:LMM}. 
In section \ref{sec:confhe}, the problem of the confined helium atom is presented. 
In section \ref{sec:sconf}, the Lagrange-mesh method in perimetric coordinates is recalled. 
A new system of coordinates is introduced in section \ref{sec:hconf} as well as 
its Lagrange-mesh implementation. 
Results for the different types of confinement are discussed on section \ref{sec:res}. 
Concluding remarks are presented in section~\ref{sec:conc}. 
\section{Principle of the Lagrange-mesh method}
\label{sec:LMM}
Let us consider $N$ mesh points $x_i$ associated with a Gauss-quadrature approximation \cite{AS65}, 
\beq
\int_0^{\infty} F(x) dx \approx \sum_{k=1}^{N} \lambda_k F(x_k).
\eeqn{1.1}
The weight coefficients $\lambda_i$ are also called Christoffel numbers. 
Lagrange functions are a set of $N$ orthonormal functions $f_j(x)$ associated 
with this mesh verifying two conditions \cite{BH86,Ba06,Ba15}. 
(i) They satisfy the Lagrange property 
\beq
f_j (x_i) = \lambda_i^{-1/2} \delta_{ij},
\eeqn{1.2}
i.e., they vanish at all mesh points, but one. 
(ii) The Gauss quadrature is exact for products of two Lagrange functions. 

Let us consider a particle of mass $m$ in a potential $V(x)$. 
This basis is used in an approximate variational calculation with the trial function 
\beq
\psi(x) = \sum_{j=1}^N c_j f_j(x).
\eeqn{1.3}
The matrix elements of potential $V(x)$ are calculated at the Gauss approximation as 
\beq
\la f_i | V | f_j \ra \approx \sum_{k=1}^N \lambda_k f_i(x_k) V(x_k) f_j(x_k) = V(x_i) \delta_{ij}
\eeqn{1.4}
because of the Lagrange property \rref{1.2}. 
The variational equations then take the form of mesh equations \cite{BH86,Ba06,Ba15} 
\beq
\sum_{j=1}^N \left( \frac{\hbar^2}{2m}\,T_{ij} + V(x_i) \delta_{ij} \right) c_j = E c_i, 
\eeqn{1.5}
where the exact or approximate matrix elements $T_{ij} = \la f_i | -d^2/dx^2| f_j \ra$ 
have simple known expressions as a function of the zeros $x_i$ and $x_j$. 
See Refs.~\cite{BH86,Ba06,Ba15} for details. 

This simple approximation can be very accurate with small numbers of mesh points 
when the Gauss approximation is valid for the potential matrix elements, 
i.e.\ when the potential and its derivatives have no singularities. 
In the presence of singularities, the basis can sometimes be regularized 
by multiplying the Lagrange functions by a convenient factor $R(x)$ 
\cite{VMB93,Ba06,Ba15}, 
\beq
\hat{f}_j (x) = \frac{R(x)}{R(x_j)} f_j (x).
\eeqn{1.6}
The regularized functions $\hat{f}_j (x)$ still verify the Lagrange conditions \rref{1.2} 
but they are not orthogonal anymore, in general. 
However, they are still orthogonal at the Gauss-quadrature approximation 
and can still be treated as orthonormal in the Lagrange-mesh method 
without significant loss of accuracy \cite{BHV02,Ba15}. 
Such a regularization is also useful when the particle is confined by an impenetrable wall 
\cite{BS08} (see section \ref{sec:hconf}). 
\section{Confined helium}
\label{sec:confhe}
We consider a two-electron atom with an infinite-mass nucleus of charge $Ze$. 
This nucleus is fixed and the electrons are characterized by 
coordinates $\ve{r}_1$ and $\ve{r}_2$ with respect to this nucleus.  
In atomic units $\hbar = m_e = a_0 = e = 1$, where $m_e$ is the electron mass 
and $a_0$ is the Bohr radius, the Hamiltonian of the helium atom reads 
\beq
H = T + V_C = -\frac{1}{2} \Delta_1 - \frac{1}{2} \Delta_2 
- \frac{Z}{r_1} - \frac{Z}{r_2} + \frac{1}{r_{12}},
\eeqn{2.1}
where 
\beq
\ve{r}_{12}=\ve{r}_1-\ve{r}_2
\eeqn{2.2}
and $\Delta_1$ and $\Delta_2$ are the Laplacians with respect to $\ve{r}_1$ and $\ve{r}_2$. 

For a free atom, the wave functions of the bound states must vanish at infinity. 
A confinement can be introduced in the problem either by forcing the wave function 
into some spherical cavity (hard confinement) or by adding a confining potential 
to $H$ (soft confinement). 

Hard confinement is obtained with the conditions 
\beq
r_1 \le R, \quad r_2 \le R.
\eeqn{2.4}
The wave function $\psi (r_1, r_2, r_{12})$ of an $S$ state must thus verify 
\beq
\psi (R, r_2, r_{12}) = \psi (r_1, R, r_{12}) = 0.
\eeqn{2.5}

Soft confinement can be obtained by adding a potential $V_{\rm conf}(r_1,r_2)$ 
which tends to a large positive constant or to infinity when $r_1$ or $r_2$ tends to infinity. 
The role of this potential is to reduce the probability density of presence 
of the electrons at large distances. 

Let us start with the soft confinement since it can be treated with the same 
code as for the free atom with only a tiny modification \cite{HB99}.
\section{Lagrange mesh for soft confinement}
\label{sec:sconf}
\subsection{Perimetric coordinates}
The system of perimetric coordinates \cite{CJ37,Pe58}
is very convenient for Lagrange-mesh calculations of three-body systems 
because the three dimensioned coordinates are independent from each other 
and vary from zero to infinity. 
Moreover, the volume element automatically regularizes 
the singularities of the three Coulomb potentials \cite{HB99}. 

The perimetric coordinates are composed of three Euler 
angles and the three coordinates 
\beq
&& x = r_1 - r_2 + r_{12}, 
\eol
&& y = -r_1 + r_2 + r_{12}, 
\eol
&& z = r_1 + r_2 - r_{12}.
\eeqn{3.3}
The volume element of the dimensioned coordinates reads
\beq
dV = (x+y) (y+z) (z+x) dx dy dz.
\eeqn{3.4}

In perimetric coordinates, the Coulomb potentials become 
\beq
V_C(x,y,z) = - \frac{2Z}{z+x} - \frac{2Z}{y+z} + \frac{2}{x+y}.
\eeqn{3.5}
With the volume element $dV$, the integrand in matrix elements 
of this potential is bounded everywhere. 
Hence the Gauss quadrature and the Lagrange-mesh method are accurate. 

The kinetic-energy operator $T$ is rather complicated \cite{Zh90}. 
It is convenient to write its matrix elements in a symmetric form \cite{GDB98,HB99}, 
\beq
\la F|T|G \ra 
= 2 \int_0^\infty dx \int_0^\infty dy \int_0^\infty dz 
\sum_{i,j=1}^3 A_{ij}(x,y,z) \frac{\partial F}{\partial x_i} \frac{\partial G}{\partial x_j},
\eeqn{3.6}
where $(x_1, x_2, x_3) \equiv (x,y,z)$. 
The coefficients $A_{ij}$ are given by 
\beq
&& A_{11} = x(y+z)(x+y+z) +x z(z+x), 
\eol 
&& A_{22} = y z(y+z) +y(z+x)(x+y+z), 
\eol
&& A_{33} = y z(y+z) + x z(z+x), 
\eol 
&& A_{12} = A_{21} = 0, 
\eol 
&& A_{13} = A_{31} = -x z(z+x), 
\eol 
&& A_{23} = A_{32} = -y z(y+z).
\eeqn{3.9}

\subsection{Lagrange mesh and functions}
The Lagrange-Laguerre functions are defined as \cite{BH86} 
\beq
f_j(x) = (-1)^j x_j^{1/2}\ \frac{L_N(x)}{x-x_j}\ e^{-x/2}, 
\eeqn{3.10}
where $L_N(x)$ is the Laguerre polynomial of degree $N$ and the 
$x_i$ are its zeros, 
\beq
L_N (x_i) = 0. 
\eeqn{3.11}
Notice that the Lagrange functions are linearly independent polynomials 
of degree $N-1$ multiplied by an exponential which is the square 
root of the Laguerre weight function $\exp(-x)$.  
The basis is thus equivalent, for example, to a basis formed 
of the Laguerre polynomials of degrees 0 to $N-1$ multiplied by $\exp(-x/2)$. 

The functions $f_j(x)$ are associated with the Gauss-Laguerre quadrature \cite{AS65}. 
They verify the Lagrange conditions \rref{1.2} 
and integrals of products of two Lagrange functions are exactly given by the Gauss quadrature 
since the integrand is the product of the Laguerre weight function $\exp(-x)$ 
by a polynomial of degree $2N-2$ \cite{Sz67}. 
Functions \rref{3.10} are exactly orthonormal over $(0,\infty)$. 

The first derivative of a one-dimensional Lagrange-Laguerre 
function at mesh points is given by 
\beq
\lambda_i^{1/2} f'_j (x_i) = (-1)^{i-j} \sqrt{\frac{x_j}{x_i}}\, \frac{1}{x_i-x_j}
\eeqn{3.14a}
for $i \ne j$ and by 
\beq
\lambda_i^{1/2} f'_i (x_i) = - \frac{1}{2x_i}.
\eeqn{3.14b}

Three-dimensional mesh and basis are obtained as follows. 
Let $x_p$ ($p = 1, \dots, N_x$), $y_q$ ($q = 1, \dots, N_y$) and $z_r$ ($r=1, \dots, N_z$) 
be the zeros of Laguerre polynomials with respective degrees $N_x$, $N_y$ and $N_z$. 
Three-dimensional Lagrange functions $F_{ijk}(x,y,z)$ associated with the mesh 
$(h_x x_p, h_y y_q, h_z z_r)$ are defined by 
\beq
F_{ijk}(x,y,z)= {\cal N}_{ijk}^{-1/2} f^{(N_x)}_i(x/h_x) f^{(N_y)}_j(y/h_y) f^{(N_z)}_k(z/h_z).
\eeqn{3.15}
The functions $f^{(N)}_i$ are given by expression (\ref{3.10}) with $N$ 
replaced by $N_x$, $N_y$ or $N_z$. 
The corresponding Christoffel numbers are denoted as 
$\lambda_i$, $\mu_j$ and $\nu_k$. 
Scale parameters $h_x$, $h_y$ and $h_z$ are introduced in order to fit 
the different meshes to the size of the actual physical problem. 
The normalization factor ${\cal N}_{ijk}$ is defined as 
\beq
{\cal N}_{ijk} = h_x h_y h_z (h_x x_i + h_y y_j) (h_x x_i + h_z z_k) (h_y y_j + h_z z_k).
\eeqn{3.16}

The Lagrange functions $F_{ijk}(x,y,z)$ satisfy the Lagrange property 
\beq
F_{ijk}(h_x x_p, h_y y_q, h_z z_r) = 
({\cal N}_{ijk} \lambda_i \mu_j \nu_k)^{-1/2} 
\delta_{ip} \delta_{jq} \delta_{kr},
\eeqn{3.17}
i.e., they vanish at all points of the three-dimensional mesh, but one. 
With the volume element \rref{3.4}, they are not orthogonal but the scalar product 
$\la F_{i' j' k'}|F_{ijk}\ra$ is calculated with the Gauss-quadrature approximation
as $\delta_{ii'} \delta_{jj'} \delta_{kk'}$. 
They are thus treated as an orthonormal basis in the method. 

The kinetic-energy matrix elements are given by 
\beq
&& \la F_{i' j' k'}|T|F_{ijk}\ra \approx 2 {\cal N}^{-1/2}_{i' j' k'} {\cal N}^{-1/2}_{ijk} h_x h_y h_z 
\eol && \times \Big\{
\delta_{jj'} \delta_{kk'}  \sum_{n} A_{11}(h_x x_n, h_y y_j, h_z z_k) 
\lambda_n f^{(N_x)\prime}_i(x_n) f^{(N_x)\prime}_{i'}(x_n) h_x^{-2} 
\eol && 
+\delta_{ii'} \delta_{kk'} \sum_{n} A_{22}(h_x x_i, h_y y_n, h_z z_k) 
\mu_n f^{(N_y)\prime}_j(y_n) f^{(N_y)\prime}_{j'}(y_n) h_y^{-2}  
\eol &&
+\delta_{ii'} \delta_{jj'} \sum_{n} A_{33}(h_x x_i, h_y y_j, h_z z_n) 
\nu_n f^{(N_z)\prime}_k(z_n) f^{(N_z)\prime}_{k'}(z_n)  h_z^{-2} 
\eol &&
+\delta_{kk'} \left[ A_{12}(h_x x_{i}, h_y y_{j'}, h_z z_k) 
(\lambda_{i} \mu_{j'})^{1/2} f^{(N_x)\prime}_{i'}(x_{i}) f^{(N_y)\prime}_{j}(y_{j'})\right. 
\eol &&
\left.+A_{12}(h_x x_{i'}, h_y y_{j}, h_z z_k) (\lambda_{i'} \mu_{j})^{1/2} 
f^{(N_x)\prime}_{i}(x_{i'}) f^{(N_y)\prime}_{j'}(y_{j})\right] (h_x h_y)^{-1} 
\eol &&
+\delta_{jj'} \left[A_{13}(h_x x_{i}, h_y y_{j}, h_z z_{k'}) (\lambda_{i} \nu_{k'})^{1/2} 
f^{(N_x)\prime}_{i'}(x_{i}) f^{(N_z)\prime}_{k}(z_{k'})\right. 
\eol &&
\left.+A_{13}(h_x x_{i'}, h_y y_{j}, h_z z_{k}) (\lambda_{i'} \nu_{k})^{1/2} 
f^{(N_x)\prime}_{i}(x_{i'}) f^{(N_z)\prime}_{k'}(z_{k})\right] (h_x h_z)^{-1} 
\eol  &&
+\delta_{ii'} \left[ A_{23}(h_x x_{i}, h_y y_{j}, h_z z_{k'}) (\mu_{j} \nu_{k'})^{1/2} 
f^{(N_y)\prime}_{j'}(y_{j}) f^{(N_z)\prime}_{k}(z_{k'})\right. 
\eol &&
\left.+A_{23}(h_x x_{i}, h_y y_{j'}, h_z z_{k})(\mu_{j'} \nu_{k})^{1/2} 
f^{(N_y)\prime}_{j}(y_{j'}) f^{(N_z)\prime}_{k'}(z_{k})\right] (h_y h_z)^{-1}  \Big\}.
\eeqn{3.23}
From now on, we consider $N_x = N_y = N$ which implies $x_i \equiv y_i$, $\lambda_i \equiv \mu_i$, and $h_x = h_y = h$. 

The Lagrange functions are used as a variational basis to expand 
an $S$-wave trial function, 
\beq
\psi(x,y,z) = \sum_{i=1}^{N} \sum_{j=1}^{i-\sigma} \sum_{k=1}^{N_z} 
C_{ijk} [2(1+\delta_{ij})]^{-1/2} [F_{ijk}(x,y,z) \pm F_{jik}(x,y,z)],
\eeqn{3.18}
where $\sigma=0$ in the symmetric case and 1 in the antisymmetric case, 
and $j \le i-\sigma$ because of the symmetry with respect to the exchange 
of electrons 1 and 2. 
The matrix representing the total potential $V = V_C + V_{\rm conf}$ 
is immediately obtained with a triple Gauss quadrature 
and the Lagrange property \rref{3.17} as 
\beq
\la F_{i'j'k'}|V|F_{ijk} \ra 
\approx V(h x_i,h x_j,h_z z_k) \delta_{ii'}\delta_{jj'}\delta_{kk'}.
\eeqn{3.20}
The variational calculation then reduces to mesh-like equations
\beq
\sum_{i=1}^{N} \sum_{j=1}^{i-\sigma} \sum_{k=1}^{N_z} \{ (1+\delta_{ij})^{-1/2} (1+\delta_{i'j'})^{-1/2} 
[\la F_{i'j'k'}|T|F_{ijk} \ra \pm \la F_{i'j'k'}|T|F_{jik} \ra] 
\eol
+ [V(h x_i, h x_j, h_z z_k) - E] 
\delta_{ii'} \delta_{jj'} \delta_{kk'} \} C_{ijk} = 0.
\eeqn{3.21}
Energies and wave functions are obtained from the eigenvalues and eigenvectors 
of a large sparse symmetric matrix. 

Mean values of a multiplicative operator $O(x,y,z)$ are simply given 
at the Gauss approximation by 
\beq
\la \psi | O(x,y,z) | \psi \ra 
\approx \dem \sum_{i=1}^{N} \sum_{j=1}^{i-\sigma} \sum_{k=1}^{N_z} 
C_{ijk}^2 [O(h x_i,h x_j,h_z z_k) + O(h x_j,h x_i,h_z z_k)].
\eeqn{3.24}
\section{Lagrange mesh for hard confinement}
\label{sec:hconf}
\subsection{Rescaled perimetric coordinates}
In an impenetrable spherical cavity of radius $R$, 
the perimetric coordinates are constrained by 
\beq
&& 0 \le x \le 2R-z, 
\eol
&& 0 \le y \le 2R-z, 
\eol
&& 0 \le z \le 2R.
\eeqn{4.1}
Hence the kinetic matrix elements \rref{3.6} must be rewritten as 
\beq
\la F|T|G \ra 
= 2 \int_0^{2R} dz \int_0^{2R-z} dx \int_0^{2R-z} dy 
\sum_{i,j=1}^3 A_{ij}(x,y,z) \frac{\partial F}{\partial x_i} \frac{\partial G}{\partial x_j}.
\eeqn{4.2}
The $z$ dependence of the upper bounds of the $x$ and $y$ coordinates is a difficulty 
for the use of the Lagrange-mesh method. 

We thus introduce a new set of coordinates $(u,v,w) \equiv (u_1,u_2,u_3)$ 
defined over $[0,1]$ by 
\beq
&& u = \frac{x}{2R-z}, 
\eol
&& v = \frac{y}{2R-z}, 
\eol
&& w = \frac{z}{2R}. 
\eeqn{4.5}
Their upper values are indeed 
\beq
&& u = 1 \quad (r_1 = R), 
\eol
&& v = 1 \quad (r_2 = R), 
\eol
&& w = 1 \quad (r_1=r_2=R, r_{12} =0).
\eeqn{4.6}
Inversely, one has  
\beq
&& x = 2Ru(1-w), 
\eol
&& y = 2Rv(1-w), 
\eol
&& z = 2Rw.
\eeqn{4.7}
The volume element then becomes  
\beq
dV = (2R)^6 (u+v) (u+w-uw) (v+w-vw) (1-w)^3 du dv dw.
\eeqn{4.8}

The Coulomb potential reads
\beq
V_C(u,v,w) = \frac{1}{R}\left[ -\frac{Z}{u+w-uw} -\frac{Z}{v+w-vw} 
+ \frac{1}{(u+v)(1-w)} \right].
\eeqn{4.24}
The integrands in its matrix elements are automatically regularized by the volume element. 
The kinetic matrix elements become 
\beq
\la F|T|G \ra 
= 2 (2R)^4 \int_0^1 du \int_0^1 dv \int_0^1 dw
\sum_{i,j=1}^3 B_{ij}(u,v,w) \frac{\partial F}{\partial u_i} \frac{\partial G}{\partial u_j}
\eeqn{4.20}
where
\beq
&& B_{11} = u(1-w) \left[ a_v b + (1-u)^2 w a_u \right], 
\eol
&& B_{22} = v(1-w) \left[ (1-v)^2 w a_v + a_u b \right], 
\eol
&& B_{33} = w(1-w)^3 \left[ v a_v + u a_u \right], 
\eol
&& B_{12} = B_{21} = uvw(1-w) \left[ (v-1) a_v + (u-1) a_u \right], 
\eol
&& B_{13} = B_{31} = uw(1-w)^2 \left[ v a_v + (u-1) a_u \right], 
\eol
&& B_{23} = B_{32} = vw(1-w)^2 \left[ (v-1) a_v + u a_u \right]
\eeqn{4.22}
with   
\beq
&& a_u = u+w-uw, 
\eol
&& a_v = v+w-vw,
\eol
&& b = u+v+w -uw-vw +uvw.
\eeqn{4.23}
\subsection{Lagrange mesh and functions}
Let us introduce a convenient basis over the $[0,1]$ interval. 
Regularized Lagrange-Legendre functions are defined by 
\beq
f_j(u) = (-1)^{N-j} \sqrt{\frac{u_j}{1-u_j}} \frac{P_N(2u-1)}{u-u_j} (1-u),
\eeqn{4.30}
where the mesh points $u_i$ are the zeros of the shifted Legendre polynomial of degree $N$, 
\beq
P_N(2u_i-1) = 0.
\eeqn{4.31}
They are associated with the Gauss-Legendre quadrature on the $[0,1]$ interval. 
The regularized Lagrange functions \rref{4.30} correspond to standard Lagrange-Legendre functions 
\cite{BH86}, shifted \cite{BHS98} and multiplied by the factor $(1-u)/(1-u_j)$, 
so that they vanish at $u = 1$. 
In \Ref{BHS98} on the contrary, they are multiplied by $u/u_j$ so that they vanish at $u = 0$. 
In \Ref{BS08}, they are multiplied by both factors simultaneously and vanish at $u = 0$ and 1. 
Notice that all these types of regularized functions 
satisfy the Lagrange conditions \rref{1.2} but are not orthogonal.  
Nevertheless they can be treated as orthonormal in Lagrange-mesh calculations 
without significant loss of accuracy \cite{BHV02,Ba15}. 

The first derivative of a Lagrange-Legendre function \rref{4.30} at mesh points is given by 
\beq
\lambda_i^{1/2} f'_j (u_i)=(-1)^{i+j} \sqrt{\frac{u_j (1-u_i)}{u_i (1-u_j)}} \frac{1}{u_i-u_j}
\eeqn{4.32}
for $i \ne j$ and by 
\beq
\lambda_i^{1/2} f'_i (u_i) = -\frac{1}{2u_i (1-u_i)}.
\eeqn{4.33}

Over a three-dimensional mesh $(u_p, v_q, w_r)$, 
where $u_p$, $v_q$, $w_r$ are solutions of \Eq{4.31} with possibly 
different values $N_u$, $N_v$, $N_w$ of $N$, 
three-dimensional Lagrange functions $F_{ijk}(u,v,w)$ are defined by 
\beq
F_{ijk}(u,v,w) = {\cal N}^{-1/2}_{i j k} f^{(N_u)}_i(u) f^{(N_v)}_j(v) f^{(N_w)}_k(w)
\eeqn{4.35}
with  
\beq
{\cal N}_{ijk}=(2 R)^6 (u_i+v_j) (u_i+w_k-u_i w_k) (v_j+w_k-v_j w_k) (1-w_k)^3.
\eeqn{4.36}
These functions satisfy the Lagrange conditions 
\beq
F_{ijk}(u_p, v_q, w_r) = ({\cal N}_{ijk} \lambda_i \mu_j \nu_k)^{-1/2} 
\delta_{ip} \delta_{jq} \delta_{kr},
\eeqn{4.37a}
where $\lambda_i$, $\mu_j$ and $\nu_k$ are now the weights of the Gauss-Legendre 
quadrature over the $[0,1]$ interval with $N_u$, $N_v$ and $N_w$ points, respectively. 
They are not orthogonal but they are orthonormal at the Gauss-quadrature approximation 
and are treated as an orthonormal basis in the method, 
\beq
\la F_{i' j' k'}|F_{ijk}\ra\ \rightarrow\ \delta_{ii'} \delta_{jj'} \delta_{kk'}.
\eeqn{4.37}
The kinetic matrix elements are given by 
\beq
&& \la F_{i' j' k'}|T|F_{ijk}\ra \approx 2 (2R)^4 {\cal N}^{-1/2}_{i' j' k'} {\cal N}^{-1/2}_{ijk} 
\eol && \times \Big\{ 
\delta_{jj'} \delta_{kk'}  \sum_{n} B_{11}(u_n,v_j,w_k) \lambda_n f^{(N_u)\prime}_i(u_n) f^{(N_u)\prime}_{i'}(u_n) 
\eol &&
+\delta_{ii'} \delta_{kk'} \sum_{n} B_{22}(u_i,v_n,w_k) \mu_n f^{(N_v)\prime}_j(v_n) f^{(N_v)\prime}_{j'}(v_n) 
\eol &&
+\delta_{ii'} \delta_{jj'} \sum_{n} B_{33}(u_i,v_j,w_n) \nu_n f^{(N_w)\prime}_k(w_n) f^{(N_w)\prime}_{k'}(w_n) 
\eol &&
+\delta_{kk'} \left[ B_{12}(u_{i},v_{j'},w_k) (\lambda_{i} \mu_{j'})^{1/2} f^{(N_u)\prime}_{i'}(u_{i}) f^{(N_v)\prime}_{j}(v_{j'})\right. 
\eol &&
\left.+B_{12}(u_{i'},v_{j},w_k) (\lambda_{i'} \mu_{j})^{1/2} f^{(N_u)\prime}_{i}(u_{i'}) f^{(N_v)\prime}_{j'}(v_{j})\right] 
\eol &&
+\delta_{jj'} \left[B_{13}(u_{i},v_{j},w_{k'}) (\lambda_{i} \nu_{k'})^{1/2} f^{(N_u)\prime}_{i'}(u_{i}) 
f^{(N_w)\prime}_{k}(w_{k'})\right. 
\eol &&
\left.+B_{13}(u_{i'},v_{j},w_{k}) (\lambda_{i'} \nu_{k})^{1/2} f^{(N_u)\prime}_{i}(u_{i'}) f^{(N_w)\prime}_{k'}(w_{k})\right] 
\eol  &&
+\delta_{ii'} \left[ B_{23}(u_{i},v_{j},w_{k'}) (\mu_{j} \nu_{k'})^{1/2} f^{(N_v)\prime}_{j'}(v_{j}) f^{(N_w)\prime}_{k}(w_{k'})\right. 
\eol &&
\left.+B_{23}(u_{i},v_{j'},w_{k})(\mu_{j'} \nu_{k})^{1/2} f^{(N_v)\prime}_{j}(v_{j'}) f^{(N_w)\prime}_{k'}(w_{k})\right] \Big\}.
\eeqn{4.40}
From now on, we take $N_u = N_v = N$. 

An $S$-wave trial function is expanded as  
\beq
\psi(u,v,w) = \sum_{i=1}^{N} \sum_{j=1}^{i-\sigma} \sum_{k=1}^{N_w} 
C_{ijk} [2(1+\delta_{ij})]^{-1/2} [F_{ijk}(u,v,w) \pm F_{jik}(u,v,w)],
\eeqn{4.38}
where $\sigma$ is defined as above. 
The potential matrix elements simply read 
\beq
&& \la F_{i' j' k'}|V_C|F_{ijk}\ra \approx \delta_{ii'} \delta_{jj'} \delta_{kk'} 
\eol
&& \times \frac{1}{R} \left[ -\frac{Z}{u_i+w_k-u_i w_k} - \frac{Z}{u_j+w_k-u_j w_k} + \frac{1}{(u_i+u_j)(1-w_k)} \right].
\eeqn{4.41}
The Lagrange-mesh equations are 
\beq
\sum_{i=1}^{N} \sum_{j=1}^{i-\sigma} \sum_{k=1}^{N_w} \{(1+\delta_{ij})^{-1/2} (1+\delta_{i'j'})^{-1/2} 
[\la F_{i'j'k'}|T|F_{ijk} \ra \pm \la F_{i'j'k'}|T|F_{jik} \ra] 
\eol
+ [V_C(u_i,u_j,w_k) - E] \delta_{ii'} \delta_{jj'} \delta_{kk'} \} C_{ijk} = 0.
\eeqn{4.39}
Notice that obtaining energies does not require calculating 
eigenvalues of a generalized eigenvalue problem 
since the basis is treated as orthonormal. 

Mean values of a multiplicative operator $O(u,v,w)$ are given 
at the Gauss approximation by 
\beq
\la \psi | O(u,v,w) | \psi \ra 
\approx \dem \sum_{i=1}^{N} \sum_{j=1}^{i-\sigma} \sum_{k=1}^{N_w} 
C_{ijk}^2 [O(u_i,u_j,w_k) + O(u_j,u_i,w_k)].
\eeqn{4.42}
\section{Results}
\label{sec:res}
All results are presented in atomic units except in Table~\ref{tab:p}. 
An error of a few units may affect the last displayed digit. 
\subsection{Soft confinement}
The study of soft confinement is performed in perimetric coordinates 
using the code of \Ref{HB99} with two modifications. 
(i) A confining potential $V_{\rm conf}$ is added to the Coulomb potential; 
this is a very simple addition in the code since no matrix elements are needed. 
(ii) The search of the lowest eigenvalues of the rather scarce Hamiltonian matrix 
is now performed with the Jacobi-Davidson algorithm \cite{BN07}. 
Mean values are calculated with \Eq{3.24} 
for the distance $r_{12}$ between electrons 
and for the distances $r_1$ and $r_2$ between the electrons and the nucleus. 

The calculations are performed to provide 12-15 significant figures 
in the fastest way, when such an accuracy can be reached with reasonable basis sizes. 
The calculation starts with a search for an optimal domain of the scale parameters $h$ and $h_z$. 
The significant digits are obtained by comparison between several calculations 
with different numbers of mesh points. 
The simplest of the calculations giving the requested accuracy is then kept. 

\begin{table}[hbt]
\caption{Convergence of the ground-state energy and the mean interparticle distances 
of a helium atom confined by the harmonic potential \rref{5.1} with $\omega = 1$ 
as a function of the numbers $N$ and $N_z$ of mesh points and the total mesh size $N_T$.}
\vspace{0.2cm}
\begin{center}
\resizebox{15.5cm}{!}{
\begin{tabular}{rrrcclll}
\hline
 $N$ & $N_z$ & $N_T$ & $h$ & $h_z$ & $E$ & $\la r_{12} \ra$ & $\la r_1 \ra = \la r_2 \ra$ \\
\hline
  10 & 10 &   550 & 0.15 & 0.20 & $-2.073\,075\,9         $   & 1.085\,817\,9            & 0.723\,715\,2            \\
     &    &       & 0.20 & 0.20 & $-2.073\,035\,387\,6    $   & 1.085\,686\,267\,07      & 0.723\,644\,406\,9       \\
  15 & 15 &  1800 & 0.15 & 0.20 & $-2.073\,035\,362\,032\,5 $ & 1.085\,685\,768\,807\,16 & 0.723\,644\,141\,796\,32 \\
     &    &       & 0.20 & 0.20 & $-2.073\,035\,362\,047\,7 $ & 1.085\,685\,768\,608\,92 & 0.723\,644\,141\,692\,94 \\
  20 & 20 &  4200 & 0.15 & 0.20 & $-2.073\,035\,362\,051\,89$ & 1.085\,685\,768\,624\,21 & 0.723\,644\,141\,700\,93 \\
     &    &       & 0.20 & 0.20 & $-2.073\,035\,362\,051\,54$ & 1.085\,685\,768\,624\,56 & 0.723\,644\,141\,701\,14 \\
  25 & 20 &  6500 & 0.15 & 0.20 & $-2.073\,035\,362\,051\,95$ & 1.085\,685\,768\,624\,25 & 0.723\,644\,141\,700\,96 \\
     &    &       & 0.20 & 0.20 & $-2.073\,035\,362\,051\,88$ & 1.085\,685\,768\,624\,25 & 0.723\,644\,141\,700\,96 \\
  25 & 25 &  8125 & 0.15 & 0.20 & $-2.073\,035\,362\,051\,94$ & 1.085\,685\,768\,624\,19 & 0.723\,644\,141\,700\,92 \\
     &    &       & 0.20 & 0.20 & $-2.073\,035\,362\,051\,86$ & 1.085\,685\,768\,623\,99 & 0.723\,644\,141\,700\,80 \\
  30 & 25 & 11625 & 0.15 & 0.20 & $-2.073\,035\,362\,051\,94$ & 1.085\,685\,768\,624\,13 & 0.723\,644\,141\,700\,88 \\
     &    &       & 0.20 & 0.20 & $-2.073\,035\,362\,051\,94$ & 1.085\,685\,768\,624\,22 & 0.723\,644\,141\,700\,93 \\
  30 & 30 & 13950 & 0.15 & 0.20 & $-2.073\,035\,362\,051\,93$ & 1.085\,685\,768\,624\,21 & 0.723\,644\,141\,700\,91 \\
     &    &       & 0.20 & 0.20 & $-2.073\,035\,362\,051\,91$ & 1.085\,685\,768\,624\,23 & 0.723\,644\,141\,700\,93 \\
\multicolumn{5}{l}{\Ref{LC09}}  & $-2.073\,035\,$             &                          &                          \\
\hline
\end{tabular}
}
\end{center}
\label{tab:OH1}
\end{table}
\begin{table}[!ht]
\caption{Ground-state energy and mean interparticle distances 
of a  helium atom confined by the harmonic potential \rref{5.1} 
as a function of the numbers $N$ and $N_z$ of mesh points 
and the scale parameters $h$ and $h_z$.}
\vspace{0.2cm}
\begin{center}
\resizebox{15.5cm}{!}{
\begin{tabular}{lrrccrll}
\hline
 $\omega$ & $N$ & $N_z$ & $h$ & $h_z$ & \multicolumn{1}{l}{$E$} & $\la r_{12} \ra$ & $\la r_1 \ra = \la r_2 \ra$ \\
\hline
0    & 30 & 25 & 0.30 & 0.35 & $-2.903\,724\,377\,034\,0$ & 1.422\,070\,255\,565\,9 & 0.929\,472\,294\,873\,7 \\  
0.01 & 30 & 25 & 0.30 & 0.35 & $-2.903\,605\,041\,422\,9$ & 1.421\,940\,016\,512\,7 & 0.929\,395\,600\,292\,0 \\  
0.05 & 30 & 25 & 0.30 & 0.35 & $-2.900\,748\,507\,134\,5$ & 1.418\,859\,663\,654\,5 & 0.927\,580\,543\,426\,0 \\  
0.1  & 25 & 25 & 0.25 & 0.30 & $-2.891\,910\,703\,103\,4$ & 1.409\,741\,024\,094\,0 & 0.922\,195\,532\,048\,2 \\  
0.25 & 25 & 25 & 0.25 & 0.25 & $-2.833\,069\,315\,397\,0$ & 1.359\,681\,803\,628\,3 & 0.892\,361\,239\,021\,3 \\  
0.5  & 25 & 20 & 0.20 & 0.20 & $-2.648\,703\,149\,419\,7$ & 1.256\,367\,735\,297\,2 & 0.829\,664\,522\,705\,9 \\  
1    & 25 & 20 & 0.20 & 0.20 & $-2.073\,035\,362\,051\,9$ & 1.085\,685\,768\,624\,2 & 0.723\,644\,141\,701\,0 \\  
2    & 20 & 20 & 0.15 & 0.15 & $-0.493\,173\,861\,504\,2$ & 0.879\,296\,076\,448\,4 & 0.592\,319\,993\,998\,0 \\
5    & 20 & 20 & 0.10 & 0.10 & $ 5.555\,021\,418\,896\,5$ & 0.619\,377\,584\,617\,5 & 0.422\,804\,512\,778\,6 \\
10   & 20 & 20 & 0.06 & 0.06 & $17.162\,191\,374\,057\,4$ & 0.459\,421\,428\,281\,5 & 0.316\,272\,749\,895\,5 \\
\hline
\end{tabular}
}
\end{center}
\label{tab:OH2}
\end{table}
As a first penetrable confinement, we choose like in \Ref{LC09} 
the harmonic potential 
\beq
V_{\rm conf} (r_1,r_2) = \dem \omega^2 (r_1^2 + r_2^2),
\eeqn{5.1}
where $\omega$ is a parameter. 
The convergence for $\omega = 1$ is studied in Table~\ref{tab:OH1} 
as a function of the numbers $N$ and $N_z$. 
The total mesh size $N_T = \dem N(N + 1) N_z$ is also given. 

The parameters $h$ and $h_z$ are roughly optimized. 
One observes a fast convergence for $h = 0.15-0.20$ and $h_z = 0.15-0.20$ 
where very accurate results are already obtained for $N = N_z = 20$. 
The results are quite insensitive to variations of $h_z$ in this interval. 
Rounding errors start to slightly deteriorate the accuracy around $N = N_z = 30$. 
For small values of $\omega$, the convergence is very similar to the free-atom case. 
See e.g.\ Table~1 of \Ref{HB99}.

\begin{table}[hbt]
\caption{Convergence of the ground-state energy and the mean interparticle distances 
of a helium atom confined by potential \rref{5.2} with $V_0 = 100$ and $R = 4$. 
The scaling parameters are $h = 0.10$ and $h_z = 0.15$.} 
\vspace{0.2cm}
\begin{center}
\begin{tabular}{rrrlll}
\hline
 $N$ & $N_z$ & $N_T$ & $E$ & $\la r_{12} \ra$ & $\la r_1 \ra = \la r_2 \ra$ \\
\hline
10 & 10 &   950 & $2.389\,517\,6           $ & 0.717\,239\,6            & 0.487\,082\,6            \\
15 & 15	&  1800 & $2.389\,521\,083\,273    $ & 0.717\,240\,933\,789     & 0.487\,083\,296\,959     \\
20 & 20	&  4200 & $2.389\,521\,083\,370\,13$ & 0.717\,240\,933\,821\,09 & 0.487\,083\,296\,966\,05 \\
25 & 25	&  8125 & $2.389\,521\,083\,370\,11$ & 0.717\,240\,933\,821\,11 & 0.487\,083\,296\,966\,08 \\
\multicolumn{3}{l}{\Ref{LC09}} & $2.389\,531$& & \\
\hline
\end{tabular}
\end{center}
\label{tab:G1}
\end{table}
The Lagrange-mesh results for various $\omega$ values are presented in Table~\ref{tab:OH2}. 
The values for the free atom ($\omega = 0$) have an accuracy of about $10^{-13}$ 
like in \Ref{HB99}. 
A relative accuracy of $10^{-12}$ is already obtained with $N = N_z = 20$. 
The confinement reduces the size of the atom when $\omega$ increases 
while increasing its energy. 
The scale factors for the weakest confinement are inspired by those of a free atom. 
When $\omega$ increases, they progressively decrease. 
To obtain a constant accuracy (thirteen significant digits), 
the numbers $N$ and $N_z$ can both decrease when the confinement becomes stonger. 
The present energies confirm the six-digit results of Ref.~\cite{LC09} for $\omega \le 1$, 
up to a possible rounding. 

\begin{table}[!hb]
\caption{Ground-state energy and mean interparticle distances 
of a helium atom confined by potential \rref{5.2} 
as a function of the numbers $N$ and $N_z$ of mesh points 
and the scale parameters $h$ and $h_z$.}
\vspace{0.2cm}
\begin{center}
\resizebox{15.5cm}{!}{
\begin{tabular}{rrrccrll}
\hline
$R$ & $N$ & $N_z$ & $h$ & $h_z$ & \multicolumn{1}{l}{$E$} & $\la r_{12} \ra$ & $\la r_1 \ra = \la r_2 \ra$ \\
\hline
\multicolumn{8}{l}{$V_0 = 25$} \\
1  & 20 & 20 & 0.10 & 0.10 & $ 9.005\,664\,919\,615\,0$ & 0.574\,015\,409\,803\,2 & 0.392\,137\,172\,502\,8 \\
2  & 20 & 20 & 0.10 & 0.10 & $ 2.205\,633\,821\,802\,6$ & 0.732\,574\,434\,299\,6 & 0.496\,887\,442\,688\,1 \\
3  & 20 & 20 & 0.10 & 0.15 & $ 0.058\,610\,162\,109\,7$ & 0.843\,146\,864\,040\,8 & 0.568\,830\,018\,099\,1 \\
4  & 20 & 20 & 0.10 & 0.15 & $-0.937\,974\,874\,207\,4$ & 0.926\,139\,670\,492\,3 & 0.622\,265\,713\,845\,8 \\
5  & 20 & 20 & 0.15 & 0.15 & $-1.493\,463\,026\,587\,6$ & 0.991\,265\,922\,032\,1 & 0.663\,853\,999\,500\,4 \\
10 & 20 & 20 & 0.20 & 0.30 & $-2.440\,318\,186\,433\,3$ & 1.180\,598\,995\,124\,2 & 0.782\,896\,512\,886\,3 \\
25 & 25 & 20 & 0.30 & 0.35 & $-2.814\,574\,813\,589\,3$ & 1.346\,698\,806\,392\,7 & 0.884\,554\,240\,409\,3 \\
100& 25 & 20 & 0.30 & 0.35 & $-2.897\,788\,906\,088\,7$ & 1.415\,741\,535\,946\,4 & 0.925\,741\,063\,258\,9 \\
\multicolumn{8}{l}{$V_0 = 100$} \\  
1  & 25 & 25 & 0.08 & 0.08 & $25.983\,328\,462\,164\,7$ & 0.407\,916\,008\,582\,9 & 0.281\,363\,150\,279\,2 \\
2  & 20 & 20 & 0.10 & 0.10 & $ 9.919\,341\,323\,720\,1$ & 0.543\,627\,679\,286\,8 & 0.372\,439\,834\,111\,5 \\
3  & 20 & 20 & 0.10 & 0.10 & $ 4.812\,589\,834\,533\,1$ & 0.640\,875\,449\,304\,4 & 0.436\,905\,248\,671\,5 \\
4  & 20 & 20 & 0.10 & 0.15 & $ 2.389\,521\,083\,370\,1$ & 0.717\,240\,933\,821\,1 & 0.487\,083\,296\,966\,1 \\
5  & 20 & 20 & 0.10 & 0.15 & $ 1.006\,687\,798\,027\,9$ & 0.780\,044\,762\,396\,9 & 0.528\,060\,835\,913\,2 \\
10 & 25 & 20 & 0.20 & 0.25 & $-1.474\,872\,423\,516\,3$ & 0.985\,908\,228\,888\,3 & 0.660\,518\,629\,618\,2 \\
25 & 25 & 20 & 0.30 & 0.35 & $-2.586\,842\,706\,183\,4$ & 1.230\,821\,212\,175\,1 & 0.813\,965\,678\,584\,9 \\
100& 25 & 20 & 0.30 & 0.35 & $-2.880\,323\,145\,511\,2$ & 1.398\,590\,911\,428\,2 & 0.915\,588\,071\,052\,0 \\
\hline
\end{tabular}
}
\end{center}
\label{tab:G2}
\end{table}
As a second potential with a penetrable confinement, we choose like in \Ref{LC09} 
\beq
V_{\rm conf} (r_1,r_2) = V_0 \left( 2 - e^{-r_1^2/R^2} - e^{-r_2^2/R^2} \right),
\eeqn{5.2}
where $V_0$ and $R$ are parameters. 

The convergence is studied in Table~\ref{tab:G1} 
for $V_0 = 100$ and $R = 4$ as a function of $N$ and $N_z$. 
Good values of the scale factors are much smaller than for the free atom \cite{HB99}. 
Here they are roughly optimized as $h = 0.1$, $h_z = 0.15$. 
The convergence is very fast. 
Good results are already obtained with $N = N_z = 10$. 
An accuracy of about 14 digits is obtained with $N = N_z = 20$ 
for the energy and for the mean distances. 

The Lagrange-mesh energies and mean distances are displayed in Table~\ref{tab:G2} 
for $V_0 = 25$ and 100 and some values of $R$. 
The scale parameters are rather small for $R = 1$ 
and must increase with $R$.
The ground-state energy is positive for the smaller $R$ values but the three-body system is nevertheless 
deeply bound with respect to the asymptotic value $2V_0$ of the confinement potential. 
Hence the wave functions decrease rather fast. 
Even for $R = 10$, the properties significantly differ from the free atom. 
The mean distances are scaled down with respect to the free atom \cite{HB99} 
but $\la r_{12} \ra$ and $\la r_1 \ra$ indicate that the confinement 
does not change much the shape of the electron distribution. 
Indeed, the ratio $\la r_{12} \ra/\la r_1 \ra$ progressively increases 
from the value 1.45 at strong confinement ($R =1$, $V_0 = 100$) to the free value 1.53. 
Our results confirm the four- and six-digit energies of \Ref{LC09} 
except for their last two digits displayed at $R = 4$ and 5. 
\subsection{Hard confinement}
Now we consider a nucleus confined in an impenetrable sphere. 
There are no scale parameters as the size is fixed by the radius $R$. 
Otherwise, the code for the new coordinate system follows the same philosophy as the previous one. 
The total mesh size is $N_T = \dem N(N + 1) N_w$ for singlet states 
and $N_T = \dem N(N - 1) N_w$ for triplet states. 
The mean values are calculated from \rref{4.42} with \rref{4.7}. 

\begin{table}[hbt]
\caption{Convergence of the ground-state energy and the mean interparticle distances 
of a helium atom confined in a sphere of radius $R$ 
as a function of the numbers $N$ and $N_w$ of mesh points and the total mesh size $N_T$.} 
\vspace{0.2cm}
\begin{center}
\resizebox{15.5cm}{!}{
\begin{tabular}{rrrlll}
\hline
 $N$ & $N_w$ & $N_T$ & $E$ & $\la r_{12} \ra$ & $\la r_1 \ra = \la r_2 \ra$ \\
\hline
\multicolumn{6}{l}{$R=0.1$} \\
 8 &  8	&  288 & $906.562\,407\,9       $ & 0.069\,580\,385\,8       & 0.049\,501\,241\,0        \\
10 & 10	&  550 & $906.562\,422\,907     $ & 0.069\,580\,382\,888\,07 & 0.049\,501\,246\,332\,6   \\
10 & 15 &  825 & $906.562\,422\,919\,887$ & 0.069\,580\,382\,884\,17 & 0.049\,501\,246\,340\,126 \\
15 & 15	& 1800 & $906.562\,429\,919\,886$ & 0.069\,580\,382\,884\,17 & 0.049\,501\,246\,340\,120 \\
15 & 20 & 2400 & $906.562\,422\,919\,888$ & 0.069\,580\,382\,884\,16 & 0.049\,501\,246\,340\,121 \\
\multicolumn{3}{l}{\Ref{MAF10}} & $906.562\,423$ & & \\
\hline
\multicolumn{6}{l}{$R=1$} \\
10 & 10 &  550 & $1.015\,754\,975\,53     $ & 0.643\,664\,253\,93     & 0.441\,796\,632\,210     \\
10 & 15 &  825 & $1.015\,754\,975\,90     $ & 0.643\,664\,253\,871\,0 & 0.441\,796\,632\,098\,55 \\
15 & 15	& 1800 & $1.015\,754\,976\,048\,33$ & 0.643\,664\,253\,878\,0 & 0.441\,796\,632\,103\,28 \\
15 & 20 & 2400 & $1.015\,754\,976\,048\,41$ & 0.643\,664\,253\,877\,9 & 0.441\,796\,632\,103\,31 \\
20 & 20	& 4200 & $1.015\,754\,976\,048\,64$ & 0.643\,664\,253\,878\,0 & 0.441\,796\,632\,103\,34 \\
20 & 25 & 5250 & $1.015\,754\,976\,048\,66$ & 0.643\,664\,253\,877\,8 & 0.441\,796\,632\,103\,41 \\
25 & 30 & 9750 & $1.015\,754\,976\,048\,67$ & 0.643\,664\,253\,878\,8 & 0.441\,796\,632\,103\,34 \\
30 & 35 &16275 & $1.015\,754\,976\,048\,54$ & 0.643\,664\,253\,876\,9 & 0.441\,796\,632\,103\,24 \\
\multicolumn{3}{l}{\Ref{AFR03}} & $1.015\,870$ & 0.643\,938 & \\
\multicolumn{3}{l}{Refs.~\cite{LC09,MAF10}} & $1.015\,755$ & & \\
\hline
\multicolumn{6}{l}{$R=10$} \\
15 & 15 & 1800 & $-2.904\,79             $ & 1.420\,767\,962\,81   & 0.928\,645\,755\,91 \\
15 & 20 & 2400 & $-2.903\,727\,2         $ & 1.422\,072\,790\,04   & 0.929\,474\,322\,57 \\
20 & 25 & 5250 & $-2.903\,724\,382\,6    $ & 1.422\,070\,196\,74   & 0.929\,472\,263\,15 \\
25 & 30 & 9750 & $-2.903\,724\,375\,625  $ & 1.422\,070\,173\,66   & 0.929\,472\,251\,58 \\
30 & 30	&13950 & $-2.903\,724\,375\,691  $ & 1.422\,070\,172\,98   & 0.929\,472\,251\,22 \\
30 & 35 &16275 & $-2.903\,724\,375\,668  $ & 1.422\,070\,172\,34   & 0.929\,472\,250\,87 \\
35 & 35	&22050 & $-2.903\,724\,375\,687  $ & 1.422\,070\,172\,96   & 0.929\,472\,251\,21 \\
\multicolumn{3}{l}{Refs.~\cite{LC09,BSM13}}& $-2.903\,724$ & & \\
\hline
\end{tabular}
}
\end{center}
\label{tab:2}
\end{table}
The convergence of Lagrange-mesh results for the ground state 
is studied in Table~\ref{tab:2} 
for various values of $R$, as a function of $N$ and $N_w$. 
For large $R$, the convergence is rather slow but it is extremely fast for small $R$. 
One obtains 15 significant figures for $R = 0.1$ and 1 but only 11 for $R = 10$. 
The situation is reversed with respect to soft-confinement calculations in perimetric coordinates 
where larger bases are needed at small $R$. 
The results are compared with Refs.~\cite{AFR03,LC09,MAF10,BSM13}. 
A good agreement with six significant digits is observed for the most recent results. 

Energies and mean distances for the ground state are gathered in Table~\ref{tab:3} 
for a number of $R$ values. 
The energies confirm the six-digit variational energies of \Ref{LC09} from 0.5 to 10 
and \Ref{BSM13} from 1 to 10. 
The present coordinate system allows a better description of smaller $R$ values. 
The earlier variational energies of \Ref{AFR03} have an absolute accuracy below 0.001. 
This reference is the only one to present $\la r_{12} \ra$ mean distances. 
Their accuracy increases from $10^{-4}$ to 0.002 when $R$ varies from 0.5 to 6. 
\clearpage
\begin{table}[hbt]
\caption{Ground-state energy and mean interparticle distances 
of a helium atom confined in a sphere of radius $R$ 
as a function of the numbers $N$ and $N_w$ of mesh points.} 
\vspace{0.2cm}
\begin{center}
\resizebox{15.5cm}{!}{
\begin{tabular}{rrrrll}
\hline
 $R$ & $N$ & $N_w$ & \multicolumn{1}{l}{$E$} & $\la r_{12} \ra$ & $\la r_1 \ra = \la r_2 \ra$ \\
\hline
   0.1 & 15 & 15 & $906.562\,422\,919\,888\ \ $ & 0.069\,580\,382\,884\,2 &  0.049\,501\,246\,340\,1 \\  
   0.2 & 15 & 20 & $206.151\,712\,932\,762\ \ $ & 0.138\,365\,789\,025\,0 &  0.097\,969\,984\,584\,9 \\  
   0.3 & 15 & 20 & $ 82.334\,517\,028\,118\ \ $ & 0.206\,218\,777\,530\,  &  0.145\,352\,111\,929\,9 \\  
   0.4 & 15 & 20 & $ 40.980\,280\,331\,705\,1$ & 0.273\,001\,759\,106\,  &  0.191\,591\,659\,047\,2 \\  
   0.5 & 15 & 20 & $ 22.741\,302\,819\,133\,5$ & 0.338\,577\,477\,653\,  &  0.236\,631\,213\,255\,9 \\  
   0.6 & 15 & 20 & $ 13.318\,127\,241\,828\,1$ & 0.402\,809\,660\,984\,  &  0.280\,412\,455\,464\,5 \\  
   0.7 & 15 & 20 & $  7.925\,216\,046\,579\,6$ & 0.465\,563\,856\,457\,  &  0.322\,876\,819\,396\,5 \\  
   0.8 & 15 & 20 & $  4.610\,407\,554\,239\,4$ & 0.526\,708\,460\,703\,  &  0.363\,966\,277\,267\,4 \\  
   0.9 & 15 & 20 & $  2.463\,235\,988\,621\,0$ & 0.586\,115\,943\,465\,  &  0.403\,624\,249\,990\,8 \\  
   1.0 & 15 & 20 & $  1.015\,754\,976\,048\,4$ & 0.643\,664\,253\,878\,  &  0.441\,796\,632\,103\,3 \\  
   2.0 & 20 & 20 & $ -2.604\,038\,275\,176\,2$ & 1.097\,202\,490\,172\,  &  0.733\,956\,380\,589\,2 \\
   3.0 & 25 & 25 & $ -2.872\,494\,886\,475\,7$ & 1.322\,925\,949\,994\,  &  0.871\,985\,220\,059\,  \\
   4.0 & 25 & 25 & $ -2.900\,485\,763\,363\,2$ & 1.399\,878\,118\,648\,  &  0.916\,962\,790\,629\,  \\
   5.0 & 30 & 30 & $ -2.903\,410\,849\,203\,1$ & 1.418\,255\,199\,620\,  &  0.927\,363\,648\,349\,  \\
   6.0 & 30 & 30 & $ -2.903\,695\,908\,930\,7$ & 1.421\,528\,373\,775\,  &  0.929\,176\,980\,774\,  \\
   7.0 & 35 & 35 & $ -2.903\,721\,911\,531\,8$ & 1.422\,002\,693\,115\,  &  0.929\,435\,862\,703\,  \\
   8.0 & 35 & 35 & $ -2.903\,724\,170\,713\,6$ & 1.422\,062\,567\,876\,  &  0.929\,468\,183\,475\,  \\
   9.0 & 35 & 35 & $ -2.903\,724\,360\,207\ \ $& 1.422\,069\,437\,714\,  &  0.929\,471\,860\,371\,  \\
  10.0 & 35 & 35 & $ -2.903\,724\,375\,687\ \ $& 1.422\,070\,172\,936\,  &  0.929\,472\,251\,212\,  \\
\hline
\end{tabular}
}
\end{center}
\label{tab:3}
\end{table}

From the radius dependence of the energies $E(R)$, 
one can deduce the pressure acting on the confined atom \cite{AFR03,LC09,BSM13}, 
\beq
P = - \frac{1}{4 \pi R^2}\,\frac{dE}{dR}.
\eeqn{5.10}
The derivative is performed numerically with a 4-point finite-difference formula. 
The results are presented in Table~\ref{tab:p}. 
As before, the significant digits are estimated by comparison 
between several calculations differing by the number of mesh points. 
Only those digits are displayed in Table~\ref{tab:p}. 
The results are presented both in atomic units and in atmospheres 
(1 a.u.\ $= 2.903\,628\,236\,775 \times 10^8$ atm). 

A high accuracy is reached up to $R = 2$. 
Beyond that value, the differences between neighboring energies become tiny 
and the relative error increases. 
The results are compared with several earlier variational calculations. 
The pressures of \Ref{AFR03} have a relative accuracy better than $10^{-4}$ 
at $R = 0.5$ and still better than $10^{-3}$ up to $R = 2$. 
They become quite poor beyond $R = 4$ as could be expected from the comparison 
between their variational calculations with different basis sizes. 
The results of \Ref{LC09} are consistent with ours for the three displayed digits 
except at $R = 5$ and 8. 
The pressures of \Ref{BSM13} have a relative accuracy varying between 0.001 and 0.006.
\begin{table}[ht]
\caption{Pressure acting on a helium atom confined in a sphere of radius $R$ 
in a.u.\ and atm. 
The powers of ten are indicated between brackets.} 
\vspace{0.2cm}
\begin{center}
\resizebox{15.5cm}{!}{
\begin{tabular}{rrrlllll}
\hline
 $R$ & $N$ & $N_w$ & $P$ (a.u.)     & $P$ (atm)        &\Ref{AFR03} &\Ref{LC09}&\Ref{BSM13}\\
\hline
   0.1 & 20 & 20 & 1.507\,426\,738\,64[5]   & 4.377\,006\,843\,19[13] &              &          &             \\
   0.2 & 20 & 20 & 4.512\,880\,064\,60[3]   & 1.310\,372\,598\,48[12] &              &          &             \\
   0.3 & 20 & 20 & 5.682\,940\,282\,5 [2]   & 1.650\,114\,587\,2 [11] &              &          &             \\
   0.4 & 20 & 20 & 1.287\,118\,285\,1 [2]   & 3.737\,312\,996\,8 [10] &              &          &             \\
   0.5 & 20 & 20 & 4.017\,215\,609\,5 [1]   & 1.166\,450\,067\,7 [10] & 1.166\,40[10]& 1.167[10]&             \\         
   0.6 & 20 & 20 & 1.534\,428\,733\,1 [1]   & 4.455\,410\,596\,9  [9] & 4.454\,36[9] & 4.455[9] &             \\
   0.7 & 20 & 20 & 6.732\,316\,374\,6       & 1.954\,814\,392\,4  [9] & 1.954\,50[9] & 1.954[9] &             \\
   0.8 & 20 & 20 & 3.266\,880\,610\,6       & 9.485\,806\,787\,1  [8] & 9.484\,10[8] & 9.485[8] &             \\
   0.9 & 20 & 20 & 1.710\,917\,738\,5       & 4.967\,869\,056\,3  [8] & 4.967\,03[8] & 4.967[8] &             \\
   1.0 & 20 & 20 & 9.510\,085\,662\,1 [-1]  & 2.761\,375\,326\,3  [8] & 2.760\,49[8] & 2.762[8] &             \\
   2.0 & 25 & 25 & 1.383\,499\,912    [-2]  & 4.017\,169\,411     [6] & 4.014\,56[6] & 4.018[6] & 4.007\,29[6]\\
   3.0 & 25 & 25 & 6.192\,831\,39     [-4]  & 1.798\,168\,01      [5] & 1.791\,50[5] & 1.798[5] & 1.793\,91[5]\\
   4.0 & 25 & 25 & 3.702\,241\,59     [-5]  & 1.074\,993\,32      [4] & 1.086\,77[4] & 1.074[4] &             \\
   5.0 & 30 & 30 & 2.365\,111\,17     [-6]  & 6.867\,403\,58      [2] & 9.021\,94[2] & 6.882[2] & 6.850\,3 [2]\\
   6.0 & 30 & 30 & 1.526\,228\,5      [-7]  & 4.431\,600\,1       [1] & 1.330\,15[2] & 4.434[2] &             \\
   7.0 & 35 & 35 & 9.871\,147         [-9]  & 2.866\,214              &              &          & 2.847\,95   \\
   8.0 & 35 & 35 & 6.399\,98          [-10] & 1.858\,31          [-1] &              & 1.868[-1]&             \\
   9.0 & 35 & 35 & 4.161\,4           [-11] & 1.208\,3           [-2] &              &          &             \\
  10.0 & 35 & 35 & 2.711\,            [-12] & 7.874\,            [-4] &              &          &             \\
\hline			        
\end{tabular}
}
\end{center}
\label{tab:p}
\end{table}

Energies and mean distances for the first excited singlet level are presented in Table~\ref{tab:4}. 
Obtaining the same accuracy as for the ground state sometimes requires higher numbers of mesh points. 
At $R = 0.1$, the excitation energy exceeds 1000 a.u. 
For $R =3$, 5, 7 and 10, a comparison is possible with \Ref{BSM13}. 
The six-digits results of \Ref{BSM13} agree with the present ones at the level of $10^{-6}$. 
The best variational energies of \Ref{FAM10} have an accuracy around one percent. 
\begin{table}[ht]
\caption{Energy and mean interparticle distances of the $2^1S$ singlet level 
of a helium atom confined in a sphere of radius $R$.} 
\vspace{0.2cm}
\begin{center}
\resizebox{15.5cm}{!}{
\begin{tabular}{rrrrll}
\hline
 $R$ & $N$ & $N_w$ & \multicolumn{1}{l}{$E$} & $\la r_{12} \ra$ & $\la r_1 \ra = \la r_2 \ra$ \\
\hline
  0.1 & 15 & 15 & $1963.757\,922\,041\,46\,\ \ \ $&  0.077\,448\,989\,286\,0 &  0.059\,004\,994\,435\,03 \\ 
  0.2 & 15 & 20 & $ 477.023\,366\,387\,052\ \ $ &  0.155\,047\,268\,447\,4 &  0.117\,679\,576\,745\,69 \\ 
  0.3 & 15 & 20 & $ 205.784\,321\,398\,280\ \ $ &  0.232\,963\,503\,164\,5 &  0.176\,011\,387\,866\,41 \\ 
  0.4 & 15 & 20 & $ 112.226\,705\,249\,011\ \ $ &  0.311\,436\,535\,890\,  &  0.233\,983\,451\,406\,40 \\ 
  0.5 & 15 & 20 & $  69.550\,421\,013\,503\ \ $ &  0.390\,828\,529\,869\,  &  0.291\,568\,743\,866\,7  \\ 
  0.6 & 15 & 20 & $  46.705\,530\,340\,508\,9$ &  0.471\,731\,678\,057\,  &  0.348\,716\,226\,254\,7  \\ 
  0.7 & 15 & 20 & $  33.131\,390\,885\,993\,4$ &  0.555\,200\,036\,019\,  &  0.405\,310\,225\,626\,3  \\ 
  0.8 & 15 & 20 & $  24.447\,862\,429\,155\,2$ &  0.643\,293\,324\,656\,  &  0.461\,031\,786\,931\,6  \\ 
  0.9 & 15 & 20 & $  18.574\,582\,880\,828\,5$ &  0.740\,350\,389\,720\,  &  0.514\,775\,812\,451\,3  \\ 
  1.0 & 15 & 20 & $  14.413\,766\,091\,552\,3$ &  0.853\,808\,811\,249\,  &  0.561\,631\,072\,086\,   \\ 
  2.0 & 20 & 25 & $   0.946\,588\,473\,886\,9$ &  1.343\,903\,122\,341\,  &  0.883\,824\,224\,122\,   \\ 
  3.0 & 25 & 25 & $  -1.114\,121\,513\,742\,1$ &  1.866\,351\,750\,214\,  &  1.203\,075\,085\,099\,   \\ 
  4.0 & 25 & 25 & $  -1.717\,517\,590\,226\,9$ &  2.408\,986\,509\,893\,  &  1.498\,595\,969\,032\,   \\ 
  5.0 & 30 & 30 & $  -1.949\,761\,321\,461\,5$ &  2.948\,630\,420\,83     &  1.782\,952\,066\,008\,   \\ 
  6.0 & 30 & 30 & $  -2.050\,702\,179\,185\,3$ &  3.435\,878\,967\,14     &  2.036\,478\,158\,792\,   \\
  7.0 & 30 & 30 & $  -2.098\,085\,074\,432\,1$ &  3.858\,190\,026\,36     &  2.254\,322\,411\,040\,   \\ 
  8.0 & 30 & 30 & $  -2.121\,511\,222\,458\,8$ &  4.215\,842\,573\,24     &  2.437\,665\,074\,483\,   \\ 
  9.0 & 35 & 35 & $  -2.133\,453\,292\,480\,3$ &  4.510\,572\,167\,17     &  2.588\,081\,892\,182\,   \\ 
 10.0 & 35 & 35 & $  -2.139\,619\,886\,178\,6$ &  4.744\,712\,449\,44     &  2.707\,196\,655\,205\,   \\ 
\hline
\end{tabular}
}
\end{center}
\label{tab:4}
\end{table}

Energies and mean distances for the lowest triplet level are gathered in Table~\ref{tab:5}. 
They are obtained from the lowest-energy solution for the spatially antisymmetric state. 
The numbers of mesh points are very close to those for the ground state. 
At $R = 0.1$, the excitation energy reaches about 1500 a.u. 
The $2^3S$ state is thus above the $2^1S$ contrary to the free-atom situation 
as already observed in \Ref{FAM10}. 
This is probably due to the additional constraint that the wave function 
must vanish for $r_1 = r_2$ in a tiny space. 
The usual ordering is recovered around $R = 1$ with an excitation energy of about 5 a.u. 
At all confinements, the triplet state is slightly less extended than the singlet excited state 
like for the free atom \cite{HB01}. 
The best variational energies of \Ref{FAM10} reach an accuracy of abour $10^{-4}$ 
above $R=1$. 
Below that value the perturbation approach is more accurate; 
its relative accuracy improves from about $10^{-3}$ at $R = 1$ to about $10^{-4}$ at $R = 0.1$. 
\begin{table}[ht]
\caption{Energy and mean interparticle distances of the $2^3S$ triplet level 
of a helium atom confined in a sphere of radius $R$.} 
\vspace{0.2cm}
\begin{center}
\resizebox{15.5cm}{!}{
\begin{tabular}{rrrrll}
\hline
 $R$ & $N$ & $N_w$ & \multicolumn{1}{l}{$E$} & $\la r_{12} \ra$ & $\la r_1 \ra = \la r_2 \ra$ \\
\hline
   0.1 & 15 & 15 & $2370.727\,023\,303\,97\ \ \ \ \,$ & 0.074\,235\,935\,806\,89& 0.049\,746\,873\,483\,20 \\
   0.2 & 15 & 15 & $ 568.187\,610\,564\,046\ \ \ $ & 0.148\,106\,984\,196\,29& 0.098\,978\,625\,497\,15 \\
   0.3 & 15 & 15 & $ 241.491\,539\,024\,040\ \ \ $ & 0.221\,580\,296\,180\,99& 0.147\,681\,956\,058\,43 \\
   0.4 & 15 & 15 & $ 129.543\,621\,160\,698\ \ \ $ & 0.294\,623\,610\,981\,39& 0.195\,843\,625\,489\,01 \\
   0.5 & 15 & 15 & $  78.820\,863\,703\,3835\ \,$ & 0.367\,205\,348\,919\,4 & 0.243\,450\,531\,053\,3  \\
   0.6 & 15 & 15 & $  51.856\,719\,933\,0814\ \,$ & 0.439\,294\,711\,654\,5 & 0.290\,489\,790\,185\,5  \\
   0.7 & 15 & 15 & $  35.951\,107\,383\,7302\ \,$ & 0.510\,861\,789\,094\,4 & 0.336\,948\,829\,834\,9  \\
   0.8 & 15 & 15 & $  25.855\,504\,026\,1363\ \,$ & 0.581\,877\,672\,009\,0 & 0.382\,815\,481\,229\,5  \\
   0.9 & 15 & 15 & $  19.089\,233\,032\,7807\ \,$ & 0.652\,314\,569\,086\,5 & 0.428\,078\,079\,126\,1  \\
   1.0 & 15 & 15 & $  14.359\,714\,920\,8699\ \,$ & 0.722\,145\,926\,846\,8 & 0.472\,725\,564\,359\,7  \\
   2.0 & 15 & 20 & $   0.560\,251\,233\,73470$ & 1.382\,427\,667\,612\,2 & 0.883\,572\,145\,681\,4  \\   
   3.0 & 20 & 25 & $  -1.370\,510\,610\,13706$ & 1.965\,771\,607\,822    & 1.228\,588\,674\,995\,6  \\   
   4.0 & 20 & 30 & $  -1.874\,611\,596\,37791$ & 2.474\,178\,173\,270    & 1.515\,265\,802\,863\,   \\   
   5.0 & 20 & 30 & $  -2.048\,044\,093\,51074$ & 2.916\,562\,529\,64     & 1.755\,588\,974\,154\,   \\   
   6.0 & 30 & 30 & $  -2.117\,816\,285\,16207$ & 3.298\,089\,015\,33     & 1.957\,801\,932\,623\,   \\
   7.0 & 30 & 30 & $  -2.148\,564\,470\,05490$ & 3.619\,012\,948\,88     & 2.125\,320\,088\,136\,   \\
   8.0 & 30 & 30 & $  -2.162\,783\,826\,65888$ & 3.878\,261\,844\,64     & 2.259\,362\,209\,113\,   \\
   9.0 & 30 & 30 & $  -2.169\,481\,052\,37618$ & 4.076\,816\,120\,79     & 2.361\,388\,479\,41      \\
  10.0 & 30 & 30 & $  -2.172\,627\,252\,24936$ & 4.219\,554\,501\,84     & 2.434\,424\,170\,24      \\
\hline
\end{tabular}
}
\end{center}
\label{tab:5}
\end{table}
\section{Concluding remarks}
\label{sec:conc}
Various types of helium confinement can be accurately treated with the Lagrange-mesh method. 
They are based on the perimetric coordinates for soft confinements or on a new coordinate 
system for the hard confinement. 
The method rapidly leads to a large but sparse symmetric matrix. 
The eigenvalue problem is not generalized, even with the new coordinate system 
for which the basis is not orthogonal. 
This striking property results from the systematic use of the Gauss quadrature 
associated with the Lagrange basis. 
This simplifying approximation does not cost accuracy \cite{BHV02,Ba15}. 

The present results improve previous works by at least five orders of magnitude 
for the energies and allow a very accurate calculation of mean interparticle distances. 
This is obtained with short computer times on a personal computer. 
The scale parameters given in the present tables allow avoiding a search for 
their optimal domain of values in future similar calculations. 
Together with the energies, approximate wave functions are also available. 
Their high accuracy is confirmed by the high accuracy on the mean values of the coordinates. 
They are available for other applications. 

A serious challenge is the extension of the method to more than three charged particles 
because a convenient coordinate system, i.e.\ where the different Coulomb singularities 
can be regularized, is not available yet. 
\section*{Acknowledgments}
This text presents research results of 
the interuniversity attraction pole programme P7/12 initiated by the Belgian-state 
Federal Services for Scientific, Technical and Cultural Affairs. 
TRIUMF receives funding via a contribution through the National Research Council Canada. 
\bibliographystyle{iopart-num}
\bibliography{confhe}
\end{document}